\documentclass[conference]{IEEEtran}
\IEEEoverridecommandlockouts
\usepackage[bookmarks=false]{hyperref}
\hypersetup{
colorlinks = true, 
hidelinks,
linkcolor=black, 
citecolor=black, 
urlcolor=black 
}

\usepackage{stfloats}
\usepackage{amsmath,amssymb,amsfonts}
\usepackage{algorithmic}
\usepackage{graphicx}
\usepackage{textcomp}
\usepackage{xcolor}
\usepackage{tikz}
\usetikzlibrary{quantikz2}
\usepackage{subcaption}
\usepackage{xspace}
\usepackage{xcolor}
\usepackage{mathtools}
\usepackage{paralist}
\usepackage{pifont}
\usepackage[most,breakable]{tcolorbox}
\usepackage{booktabs}
\usepackage{xspace}
\usepackage{bbding}
\usepackage{colortbl}
\usepackage{tabularx}
\usepackage{siunitx}
\usepackage{nicefrac}
\usepackage[inline]{enumitem}
\usepackage{quantikz}
\usepackage{amsbsy}

\makeatletter 
\newcommand{\linebreakand}{%
\end{@IEEEauthorhalign}
\hfill\mbox{}\par
\mbox{}\hfill\begin{@IEEEauthorhalign}
}
\makeatother 

\newcommand{\averageAccuracy}{\ensuremath{\mathit{avgAcc}}\xspace}
\newcommand{\ourApproach}{\ensuremath{\mathtt{Qlinical}}\xspace}
\newcommand{\ourApproachComb}[1]{\ensuremath{\mathtt{Qlinical}_{#1}}\xspace}
\newcommand{\caress}{\texttt{CaReSS}\xspace}
\newcommand{\guri}{\texttt{GURI}\xspace}
\newcommand{\evomaster}{\texttt{EvoMaster}\xspace}
\newcommand{\evoclass}{\texttt{EvoClass}\xspace}
\newcommand{\comb}{\ensuremath{\mathtt{comb}}\xspace}
\newcommand{\hparm}{\ensuremath{\mathtt{hp}}\xspace}
\newcommand{\bestComb}{\ensuremath{\mathtt{bestComb}}\xspace}
\newcommand{\bestCombTwo}{\ensuremath{\mathtt{bestComb_{2hps}}}\xspace}
\newcommand{\bestCombThree}{\ensuremath{\mathtt{bestComb_{3hps}}}\xspace}
\newcommand{\combTuple}[4]{$\langle #1,$ $#2,$ $#3,$ $#4\rangle$\xspace}
\newcommand{\combTupleThree}[3]{$\langle #1,$ $#2,$ $#3, -\rangle$\xspace}
\newcommand{\combTupleTwo}[2]{$\langle #1,$ $#2, -, -\rangle$\xspace}
\newcommand{\featNum}{\ensuremath{\mathtt{num\_feat}}\xspace}
\newcommand{\repNum}{\ensuremath{\mathtt{num\_rep}}\xspace}
\newcommand{\entalg}{\ensuremath{\mathtt{entagl}}\xspace}
\newcommand{\ansatz}{\ensuremath{\mathtt{ansatz}}\xspace}

\newcommand{\entalgFL}{\ensuremath{\mathtt{fl}}\xspace}
\newcommand{\entalgLN}{\ensuremath{\mathtt{ln}}\xspace}
\newcommand{\entalgRL}{\ensuremath{\mathtt{rl}}\xspace}
\newcommand{\entalgPW}{\ensuremath{\mathtt{pw}}\xspace}
\newcommand{\entalgCL}{\ensuremath{\mathtt{cl}}\xspace}
\newcommand{\entalgSCA}{\ensuremath{\mathtt{sca}}\xspace}
\newcommand{\ansatzRA}{\ensuremath{\mathtt{ra}}\xspace}
\newcommand{\ansatzES}{\ensuremath{\mathtt{es}}\xspace}
\newcommand{\ansatzEPFS}{\ensuremath{\mathtt{ep\_fs}}\xspace}
\newcommand{\ansatzEPIS}{\ensuremath{\mathtt{ep\_is}}\xspace}
\newcommand{\ansatzPTD}{\ensuremath{\mathtt{ptd}}\xspace}

\def\BibTeX{{\rm B\kern-.05em{\sc i\kern-.025em b}\kern-.08em
T\kern-.1667em\lower.7ex\hbox{E}\kern-.125emX}}

\begin{document}

\title{Quantum Neural Network Classifier for Cancer Registry System Testing: A Feasibility Study\thanks{The work is supported by the Qu-Test project (Project \#299827) funded by the Research Council of Norway and Simula's internal strategic project on quantum software engineering. S. Ali is also supported by Oslo Metropolitan University's Quantum Hub. P. Arcaini is supported by Engineerable AI Techniques for Practical Applications of High-Quality Machine Learning-based Systems Project (Grant Number JPMJMI20B8), JST-Mirai.}}

\author{
\IEEEauthorblockN{Xinyi Wang}
\IEEEauthorblockA{\textit{Simula Research Laboratory}\\
\textit{University of Oslo}\\
Oslo, Norway \\
xinyi@simula.no}
\and
\IEEEauthorblockN{Shaukat Ali}
\IEEEauthorblockA{\textit{Simula Research Laboratory} \\
\textit{and Oslo Metropolitan University}\\
Oslo, Norway \\
shaukat@simula.no}
\and
\IEEEauthorblockN{Paolo Arcaini}
\IEEEauthorblockA{\textit{National Institute of Informatics} \\
Tokyo, Japan \\
arcaini@nii.ac.jp}
\linebreakand
\IEEEauthorblockN{Narasimha Raghavan Veeraragavan}
\IEEEauthorblockA{\textit{Cancer Registry of Norway,}\\
\textit{Norwegian Institute of Public Health,}\\ 
\textit{Oslo, Norway}\\
nara@kreftregisteret.no}
\and
\IEEEauthorblockN{Jan~F.~Nygård}
\IEEEauthorblockA{\textit{Cancer Registry of Norway,}\\
\textit{Oslo, Norway} \\
\textit{and The Arctic University of Norway}\\
Tromsø, Norway \\
jfn@kreftregisteret.no}
}

\maketitle

\begin{abstract}
The Cancer Registry of Norway (CRN) is a part of the Norwegian Institute of Public Health (NIPH) and is tasked with producing statistics on cancer among the Norwegian population. For this task, CRN develops, tests, and evolves a software system called Cancer Registration Support System (\caress). It is a complex socio-technical software system that interacts with many entities (e.g., hospitals, medical laboratories, and other patient registries) to achieve its task. For cost-effective testing of \caress, CRN has employed \evomaster, an AI-based REST API testing tool combined with an integrated classical machine learning model. Within this context, we propose \ourApproach to investigate the feasibility of using, inside \evomaster, a Quantum Neural Network (QNN) classifier, i.e., a quantum machine learning model, instead of the existing classical machine learning model. Results indicate that \ourApproach can achieve performance comparable to that of \evoclass. We further explore the effects of various QNN configurations on performance and offer recommendations for optimal QNN settings for future QNN developers.
\end{abstract}



\section{Introduction}
As machine learning (ML) and quantum computing (QC) advance rapidly, there is an increased interest in enhancing classical ML with QC. Such interest has resulted in the area of quantum machine learning (QML). QML has a new class of ML models that leverages the unique properties of quantum mechanics, such as superposition and entanglement, to train ML models on quantum computers. Research has demonstrated the advantages of QML for solving real-world problems, such as image processing, natural language processing, and pattern recognition~\cite{zhou2018quantum, lorenz2023qnlp, das2023quantum}. Some commercial-industrial applications based on QML have also been developed~\cite{bayerstadler2021industry,bova2021commercial}.


As the QML field is establishing, assessing how various QML algorithms perform in real-world contexts is essential. To this end, we explore the application of one such algorithm, i.e., Quantum Neural Network (QNN)~\cite{kwak2021quantum}, in the context of the Cancer Registry of Norway (CRN), a division in the Norwegian Institute of Public Health (NIPH)---a governmental body responsible for various activities related to Norwegian citizens' health. CRN collects data, analyzes it, and provides statistics to the public on cancer incidences in Norway for conducting research. To facilitate this, it built the Cancer Registration Support System (\caress), a complex socio-technical system responsible for receiving data, processing it, and finally producing the processed data and statistics for various purposes (e.g., for governmental agencies and researchers). The software system receives data as cancer messages of patients from different medical entities (e.g., medical labs and hospitals) in Norway and registers this patient information. These received messages are then validated by \guri, an automated key component within \caress responsible for handling the received data and validating it against medical rules, which are implemented based on various information, such as medical knowledge, standards, and regulations.

\caress undergoes continuous evolution due to factors such as the emergence of new medical standards, and software updates, which lead to frequent changes in validation rules in \guri. Thus, \guri requires automated testing tools for continuous testing. \evomaster~\cite{arcuri2018evomaster,arcuri2021evomaster}, an automated test case generation tool aiming at Representational State Transfer (REST) Application Programming Interfaces (APIs), is used for testing \guri. It implements evolution algorithms to generate various requests, which are cancer messages, to identify faults of \guri effectively. However, there is the possibility that \evomaster generates \emph{invalid} requests that are against the requirements of specific REST endpoints (i.e., validation endpoint and aggregation endpoint). When testing \guri with \evomaster, each request should be sent to \guri running in real-time. Executing those invalid requests leads to unnecessary \guri execution costs and negatively affects the performance of \caress during operation. However, identifying all potential invalid combinations or patterns within the requests is complex. Thus, \evoclass~\cite{isaku2023cost} has been proposed to use a machine learning model to enhance the performance of \evomaster in this context. It is a classifier that predicts whether an API request generated by the \evomaster testing tool is likely to be invalid. It filters out those requests before sending them to \guri, thereby reducing the number of requests executed during testing.

Motivated by the current active investigation of QML in practice in various domains, we propose \ourApproach to investigate the feasibility of applying QNN, a specific type of QML model, as an alternative to \evoclass, assessing its potential in the context of a real-world application. We evaluate the performance of QNN models across various configurations, and results show that QNN models can achieve performance comparable to that of the classical machine learning approach \evoclass, but by using significantly fewer training features in certain configurations. Then, we examine the impact of four hyperparameters and their interactions on QNN performance. Finally, we offer configuration recommendations for the four hyperparameters to assist future QNN developers.

\section{Background}\label{sec:background}
\subsection{Quantum Computing}
\begin{figure}[!tb]
\centering
\includegraphics[width=0.3\linewidth]{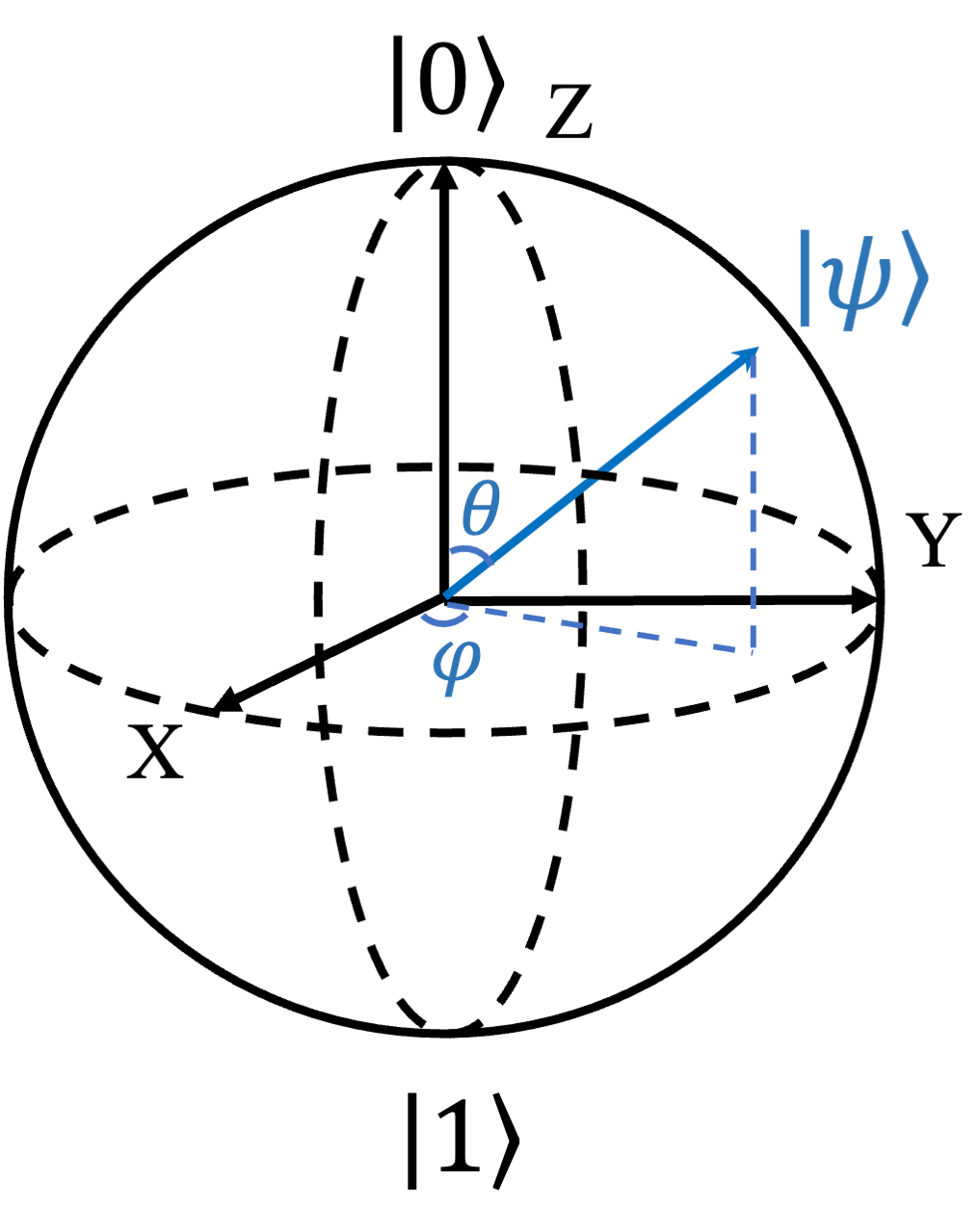}
\caption{Qubit visualization as Bloch sphere}
\label{fig:qubit}
\end{figure}

Traditional computers operate on bits, which can be either 0 or 1. In contrast, Quantum Computing (QC) utilizes \emph{quantum bits} (\emph{qubits}) for computations, which can be in a superposition of both $\ket{0}$ and $\ket{1}$ states at the same time with different probabilities. The superposition is one of the key principles that enables QC to perform complex calculations more efficiently. Another key principle in QC is \emph{entanglement}, where involved qubits are strongly correlated and the operation on one qubit will simultaneously affect the other qubit. Commonly, to visualize a qubit's state, a {\it Bloch sphere} (see Fig.~\ref{fig:qubit}) is used.
Any position in the sphere corresponds to a \emph{quantum state} of the qubit. Thus, a qubit's state can be described through the rotation angles relative to the $x$, $y$, and $z$ axes.

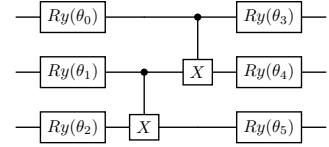
\begin{figure}[!tb]
\centering
\resizebox{0.5\columnwidth}{!}{
\begin{quantikz}
&\gate{Ry(\theta_0)}&&\ctrl{1}&\gate{Ry(\theta_3)}&\\
&\gate{Ry(\theta_1)}&\ctrl{1}&\gate{X}& \gate{Ry(\theta_4)}& \\
&\gate{Ry(\theta_2)}&\gate{X}&&\gate{Ry(\theta_5)}&
\end{quantikz}
}
\caption{An example of quantum circuit}
\label{fig:ansatz}
\end{figure}

In this work, we use \emph{gate-based QC} to build \emph{quantum neural networks}, which are built upon \emph{quantum circuits}. Fig.~\ref{fig:ansatz} shows an example of a quantum circuit example of three qubits.
Each line represents a qubit. By applying different quantum gates to those qubits, we can rotate the qubit around three axes and change their states to perform computation.

Below, we describe the main quantum gates used in this work to build a QNN circuit, including some \emph{one-qubit gates} and \emph{two-qubit gates}. Notably, the two-qubit gates play an important role in creating entanglement between the two involved qubits.
\begin{compactitem}
\item \textbf{Hadamard gate ($H$)}: A single-qubit gate that puts a qubit into an equal superposition of gates, resulting in equal probability of measuring the qubit as $\ket{0}$ or $\ket{1}$.
\item \textbf{Phase gate ($P$)}: A single-qubit gate that applies a partial $z$-rotation to the $\ket{1}$ state with customized radians.
\item \textbf{Pauli gates ($X/Y/Z$)}: A set of single-qubit gates rotating around the $x/y/z$-axis through $\pi$ radians.
\item \textbf{Standard rotation gates ($R_X/R_Y/R_Z$)}: A set of single-qubit gates rotating around the $x/y/z$ axis through a customized radians.
\item \textbf{Controlled Pauli gates ($C_X/C_Y/C_Z$)}: A set of two-qubit gates involving a control and a target qubit. It applies the Pauli-$X/Y/Z$ gate to the target qubit if the control is in state $\ket{1}$.
\item \textbf{Controlled rotation gates ($C_{R_X}/C_{R_Y}/C_{R_Z}$)}: A set of two-qubit gates involving a control and a target qubit. It applies the $R_X/R_Y/R_Z$ gate with a customized radian on the target qubit if the control qubit is in state $\ket{1}$.
\item \textbf{Two qubit rotation gates ($R_{XX}/R_{YY}/R_{ZZ}$)}: A set of two-qubit gates rotating the two qubits around $x/y/z$ axes simultaneously with a customized radians.
\end{compactitem}

At the end of a quantum circuit, a set of \emph{observables} is applied to different qubits to measure the final quantum state. This causes the measured qubits to collapse and is key in extracting outputs from the quantum circuits. Common observables include \emph{Pauli operators}, $X$, $Y$ and $Z$, which can obtain partial information of $x$, $y$, $z$ bases from the measured qubits, respectively.

\subsection{Quantum Neural Network}\label{sec:qnn}
QNN is a variational quantum algorithm containing parameterized quantum circuits. The following key components are part of a QNN circuit.

\subsubsection{Feature map}
\begin{figure}[!tb]
\centering
\includegraphics[width=0.7\linewidth]{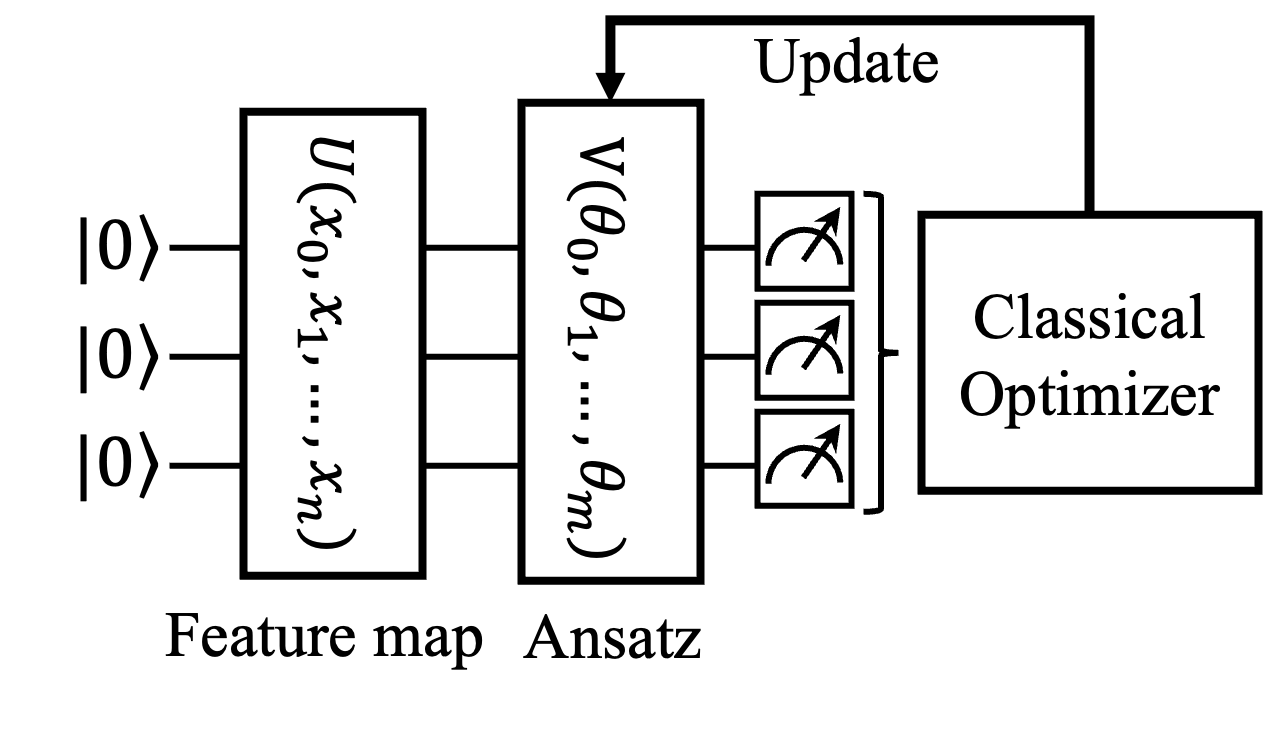}
\caption{QNN structure}
\label{fig:qnn_structure}
\end{figure}

In a QNN, classical input information is commonly first encoded into quantum space with a {\it feature map} as shown in Fig.~\ref{fig:qnn_structure} to enable processing by the following quantum operations in the ansatz.
These input feature values are translated into parameters of quantum circuits to create a quantum state. This process is called {\it quantum embedding}. There have been a few widely used embedding techniques, such as angle embedding, which encodes each feature value of input into rotation angles of the qubits.

\subsubsection{Ansatz}
Once the input features are encoded into a quantum state, an \emph{ansatz} (shown in Fig.~\ref{fig:qnn_structure}), typically with multiple \emph{repetitions}, is applied with trainable parameters, similar to classical neural networks. Those parameters are usually rotation angles of quantum gates. The structure of the ansatz is fixed before training. Thus, the choice of ansatz is significant since, when training starts, only the parameters (i.e., rotation angles) will be optimized for the specific task, such as regression or classification problems. The optimization process is performed based on a cost function.

\subsubsection{Observables}
A list of observables to extract partial information of the circuit outputs from the three axes.

\subsubsection{Optimizer}
With a defined cost function, the classical optimizer aims to minimize the loss based on the outputs from the circuit. In each iteration, the classical optimizer updates all trainable parameters in the ansatz until meeting the stopping condition.

The details of the QNN circuits used in this paper will be explained in detail in Sect.~\ref{sec:methodology}.

\section{Industrial Context}\label{sec:industrialContext}
The Cancer Registry of Norway (CRN) collects data about cancer patients, such as basic information (e.g., gender and age), as well as cancer information of cancer types and treatment records, and provides the data and statistics to various stakeholders, such as researchers, doctors, and policymakers. An interactive, human-in-the-loop decision support system, Cancer Registration Support System (\caress), is developed to collect, validate, and process patient information (i.e., cancer messages) from medical entities such as hospitals, clinical departments, and other registries.

\guri is a key component inside \caress, that is responsible for receiving cancer messages via a web application through REST APIs and performing validation and aggregation on them with hundreds of medical rules. Such medical rules are built based on diverse information, such as the domain knowledge from medical experts, standards, and regulations. However, these rules change constantly as CRN keeps introducing new rules and revising existing ones due to, e.g., changes in regulations and standards as well as when new scientific knowledge becomes available. Additionally, software updates in \caress are ongoing on a regular basis, which results in changing the software processing these rules. Thus, \caress (specifically \guri) requires continuous evolution and must be tested continuously with automated testing tools to ensure data quality, accuracy, and security, which is crucial since \caress supports critical decision-making by policymakers and epidemiological studies by researchers~\cite{laaber2023challenges}.

\evomaster (i.e., an AI-based system-level testing tool focusing on testing REST APIs~\cite{arcuri2019restful}) is used to test \guri by automatically generating HTTP requests (i.e., including cancer messages inside) based on the OpenAPI Specification (OAS) Swagger. The testing focuses on two REST endpoints: the {\it validation endpoint}, validating cancer messages with validation rules, and the {\it aggregation endpoint}, consolidating cancer messages into cancer cases with aggregation rules. 

An HTTP request can effectively test the core functionality of the two endpoints only if it is {\it meaningful}, i.e., it meets the specific input requirements for each endpoint, such as correct data types and constraints.
We refer to these meaningful requests as {\it valid} requests; otherwise, they are termed {\it invalid} requests. The validity of a request cannot be determined until it is executed in \guri. Specifically, once a request is generated, \guri executes it and returns a response. The request is ultimately deemed valid only if the status code in the response is ``200'', indicating an ``OK'' server response. Any other status code signifies that the request is {\it invalid}. Executing invalid requests can unnecessarily increase testing costs, as \guri processes these requests in real-time, potentially impacting the system's operational performance. 

However, there is a high possibility that \evomaster generates {\it invalid} requests, and it is difficulty to identify them due to the complexity of determining the request validity based on query, path, and body parameters/content. To avoid unnecessary testing costs in \guri, \evoclass~\cite{isaku2023cost} was introduced to filter out those requests that are highly likely to be {\it invalid} before actually executing them in \guri. \evoclass uses a machine-learning classifier and is trained based on requests generated by \evomaster and their corresponding responses with the status codes.

Motivated by current research on real-world applications of QML in various domains~\cite{gujju2024quantum}, this work proposes to explore the feasibility of applying QML, QNNs in particular, to aid the testing of the \guri system. Instead of using classical machine learning approaches, we train the QNN classifiers to identify possibly invalid requests generated by \evomaster. Even though QNNs are quite an early-stage technology, with their full potential still limited by the current small-scale quantum computers and the noise in the computations, it is still worth investigating their feasibility in real-world contexts.

\section{Methodology}\label{sec:methodology}

\subsection{Overview}\label{sec:overview}
We propose \ourApproach, a QNN-based method, to classify requests generated by \evomaster into two categories, {\it valid} and {\it invalid}. This approach aims to prevent the execution of invalid requests in \guri during the testing process, thereby reducing testing costs.

\begin{figure}[!tb]
\centering
\includegraphics[width=0.99\linewidth]{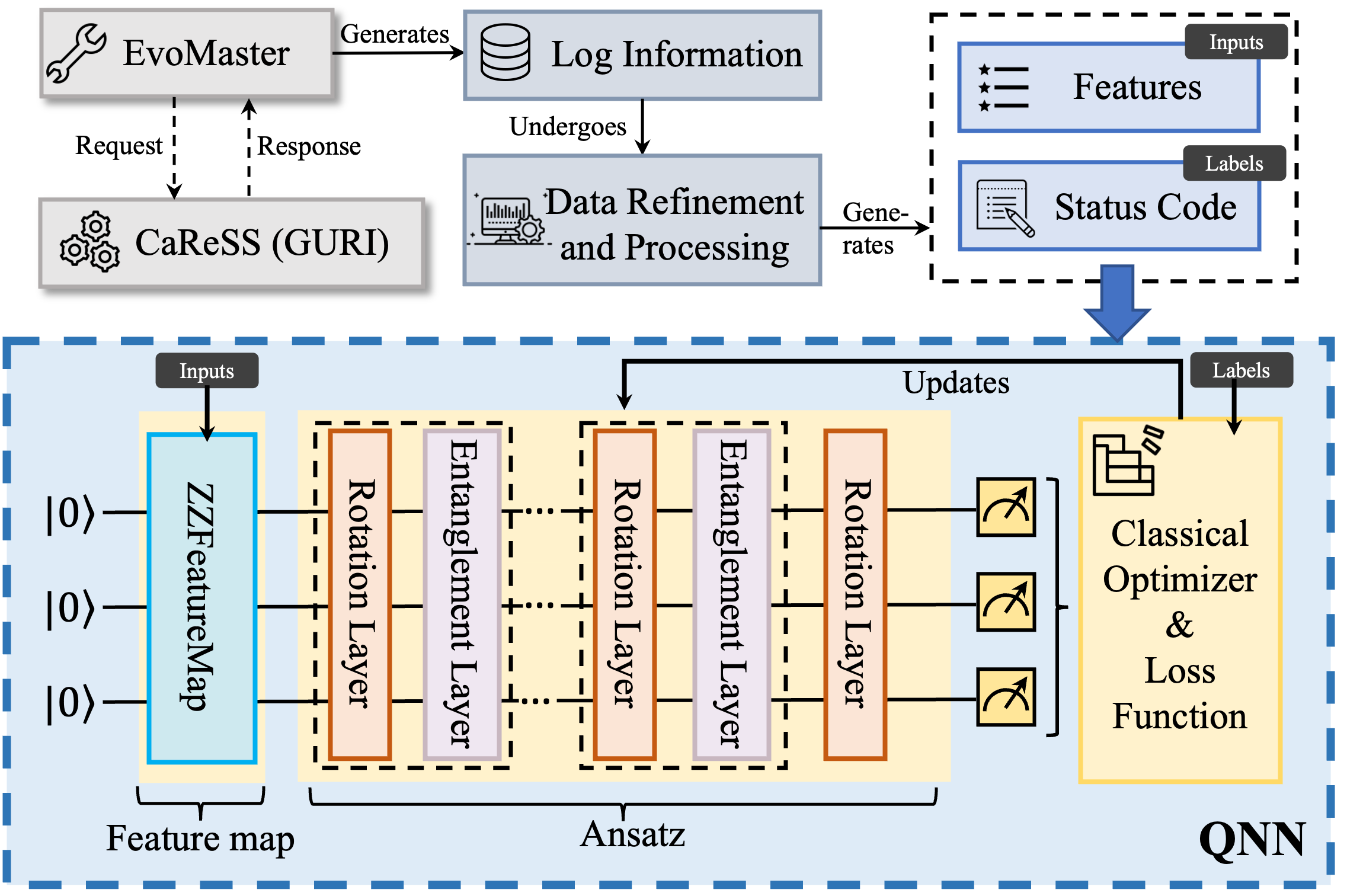}
\caption{Training process of \ourApproach}
\label{fig:overview}
\end{figure}

Fig.~\ref{fig:overview} illustrates the training process of \ourApproach.
\evomaster begins by generating a set of requests based on the OAS schema, which are then sent to \guri for execution. \guri processes these requests and returns the corresponding responses. Both the requests and their responses are logged into a file for further analysis. Then, useful data is extracted and processed from the log files. 
For each request, the corresponding generated status code in the response is considered as label for training, where ``200'' represents {\it valid} while other codes are considered {\it invalid}. Other information related to the generation information is converted into feature values suitable for training, which are inputs of the QNN model. Suppose there are $n$ features in the feature set $\{F_0,$ $F_1,$ $\ldots,$ $F_{n-1}\}$, the feature values are represented as $\boldsymbol{v}=\{v_0,$ $v_1,$ $\ldots,$ $v_{n-1}\}$.

Next, in the training process, for each request, the corresponding feature values are encoded into the QNN model. As explained in Sect.~\ref{sec:qnn}, we encode feature values into a \textit{feature map}. Specifically, in \ourApproach, we use the \textit{ZZ feature map}~\cite{havlivcek2019supervised} to encode the classical values into quantum space. Then, an ansatz is applied to process the inputs, which contain parameterized quantum gates for training. This work mainly implements two-local circuits as ansatz~\cite{bharti2022noisy}, typically consisting of alternating rotation layers and entanglement layers. Specifically, the gates applied in blocks of rotation layers and entanglement layers can be customized. We can also specify the number of repetitions of these alternating rotation layers and entanglement layers, denoted as \repNum, which decides the complexity of the circuit. At the end of the circuit execution, observables are applied to all the qubits to measure the output results. We apply Pauli-$Z$ operator on each qubit as observables to extract output information. A cost function is defined to calculate the distance between the current output and the target output (i.e., the label). A classical optimizer is applied to minimize the cost function and update the parameters of quantum gates in the ansatz in each iteration.


\subsection{QNN Circuits}
In this paper, we perform an empirical study on the performance of a specific quantum feature map combined with various ansatz. Both the feature map and the ansatz implemented in this paper belong to the two-local circuit category, which are parameterized circuits where interactions are limited to pairs of qubits at a time, rather than involving multiple qubits in a single operation. These circuits are commonly used in quantum machine learning and are more practical to implement on quantum hardware, as they only require single- and two-qubit gates. Specifically, these circuits comprise multiple blocks of rotation and entanglement layers. The rotation layers consist of single-qubit rotation gates, while entanglement layers usually consist of two-qubit gates. Here we first introduce the entanglement structure, as they are applied in both the feature map and the ansatz.

\subsubsection{Entanglement structure}\label{sec:entanglementstructure}
The structure of entanglement layers can be customized. Here, we first introduce the six entanglement structures that we consider. Note that the specific gates used depend on the applied ansatz types or feature map, and they are not decided by entanglement itself.
\begin{compactitem}
\item \textit{Linear entanglement (\entalgLN):} Each qubit $q_i$ is entangled with $q_{i+1}$. This means that there is at least one two-qubit gate applied between every pair of $q_i$ and $q_{i+1}$. Taking the entanglement layer in Fig.~\ref{fig:zzfeaturemap} as an example (i.e., right side of the red dashed line), \textit{ZZ feature map} uses three gates (i.e., two $C_X$ gates and a Phase gate in the blue box as an entanglement unit) to entangle and rotate the corresponding two qubits.
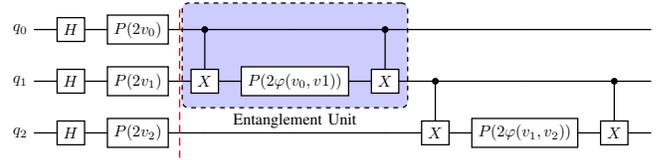
\begin{figure}[!tb]
\centering
\resizebox{0.99\columnwidth}{!}{
\begin{quantikz}
\lstick{$q_0$}& \gate{H} & \gate{P(2v_0)}\slice{} & \ctrl{1}\gategroup[2,steps=3,style={dashed,rounded
corners,fill=blue!20, inner
xsep=2pt},background,label style={label
position=below,anchor=north,yshift=-0.2cm}]{{Entanglement Unit}}  & & \ctrl{1} & & &&\\
\lstick{$q_1$}& \gate{H} & \gate{P(2v_1)} & \gate{X} & \gate{P(2\varphi(v_0, v1))} & \gate{X} & \ctrl{1} & & \ctrl{1} &\\
\lstick{$q_2$}& \gate{H} & \gate{P(2v_2)} & & & & \gate{X} & \gate{P(2\varphi(v_1, v_2))} & \gate{X}&
\end{quantikz}
}
\caption{An example of \textit{ZZ feature map} with \textit{linear entanglement} (\entalgLN)}
\label{fig:zzfeaturemap}
\end{figure}
$q_0$ and $q_1$ are entangled, as well as $q_1$ and $q_2$. However, $q_0$ and $q_2$ are not directly entangled.
\item \textit{Full entanglement (\entalgFL):} Every two qubits are entangled, meaning there is at least one two-qubit gate applied between every two qubits. Taking the entanglement layer in Fig.~\ref{fig:realfull} as an example (i.e., between red dashed lines $a$ and $b$), $C_X$ gates are applied between qubit $q_0$ and $q_1$, $q_0$ and $q_2$, as well as $q_1$ and $q_2$ to entangle all three qubits.
\begin{figure}[!tb]
\centering
\centering
\resizebox{0.65\columnwidth}{!}{
\begin{quantikz}
\lstick{$q_0$} & \gate{R_y(\theta_0)} \slice{a} & \ctrl{1} & \ctrl{2} & \slice{b} & \gate{R_y(\theta_3)} & \\
\lstick{$q_1$} & \gate{R_y(\theta_1)} & \gate{X} & &\ctrl{1}& \gate{R_y(\theta_4)} & \\
\lstick{$q_2$} & \gate{R_y(\theta_2)} & & \gate{X} &\gate{X}& \gate{R_y(\theta_5)} & 
\end{quantikz}
}
\caption{\textit{Real amplitudes} circuit (\ansatzRA) with \textit{full entanglement} (\entalgFL)}
\label{fig:realfull}
\end{figure}
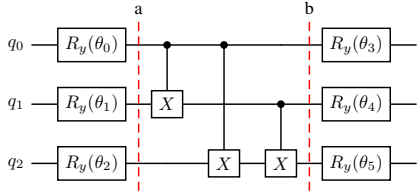
\item \textit{Pairwise entanglement (\entalgPW):} It consists of two layers. In the first layer, qubits with even numbers are entangled with the next qubit. In the second layer, qubits with odd numbers are entangled with the next qubit. Taking the entanglement layer in Fig.~\ref{fig:paulipair} as an example (i.e., between red dashed lines $b$ and $c$), $C_Z$ gates are first applied between qubit $q_0$ and $q_1$, $q_2$ and $q_3$, since 0 and 2 are even numbers.
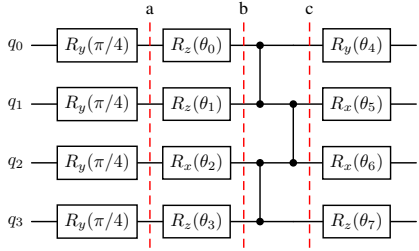
\begin{figure}[!tb]
\centering
\resizebox{0.65\columnwidth}{!}{
\begin{quantikz}
\lstick{$q_0$} & \gate{R_y(\pi/4)} \slice{a} & \gate{R_z(\theta_0)} \slice{b} & \ctrl{1} & \slice{c} & \gate{R_y(\theta_4)} & \\
\lstick{$q_1$} & \gate{R_y(\pi/4)} & \gate{R_z(\theta_1)} & \control{} & \ctrl{1} & \gate{R_x(\theta_5)} & \\
\lstick{$q_2$} & \gate{R_y(\pi/4)} & \gate{R_x(\theta_2)} & \ctrl{1} &\control{} & \gate{R_x(\theta_6)} & \\
\lstick{$q_3$} & \gate{R_y(\pi/4)} & \gate{R_z(\theta_3)} & \control{} & & \gate{R_z(\theta_7)} & 
\end{quantikz}
}
\caption{\textit{Pauli two-design} circuit (\ansatzPTD) with \textit{pairwise entanglement} (\entalgPW)}
\label{fig:paulipair}
\end{figure}
In the second layer, $q_1$ and $q_2$ are entangled with a $C_Z$ gate.
\item \textit{Reserve linear entanglement (\entalgRL):} Gates are the same as linear entanglement except in a reserved order. Taking the entanglement layer in Fig.~\ref{fig:excitationreiswap} as an example (i.e., between red dashed lines $a$ and $b$), \textit{excitation preserving} circuit applies $R_{XX}$ and $R_{YY}$ gates as a entanglement unit to entangle qubits.
\begin{figure}[!tb]
\centering
\resizebox{0.99\columnwidth}{!}{
\begin{quantikz}
\lstick{$q_0$} & \gate{R_z(\theta_0)} \slice{a} &&& \gate[2]{R_{xx}(\theta_4)} & \gate[2]{R_{yy}(\theta_4)} \slice{b} & \gate{R_z(\theta_5)}& \\
\lstick{$q_1$} & \gate{R_z(\theta_1)} &\gate[2]{R_{xx}(\theta_3)}\gategroup[2,steps=2,style={dashed,rounded
corners,fill=blue!20, inner
xsep=2pt},background,label style={label
position=below,anchor=north,yshift=-0.2cm}]{{Entanglement Unit}}&\gate[2]{R_{yy}(\theta_3)}& & & \gate{R_z(\theta_6)}& \\
\lstick{$q_2$} & \gate{R_z(\theta_2)} &&& & & \gate{R_z(\theta_7)}&
\end{quantikz}
}
\caption{\textit{Excitation preserving} circuit for mode \textit{``iswap''} (\ansatzEPIS) with \textit{reserve-linear entanglement} (\entalgRL)}
\label{fig:excitationreiswap}
\end{figure}
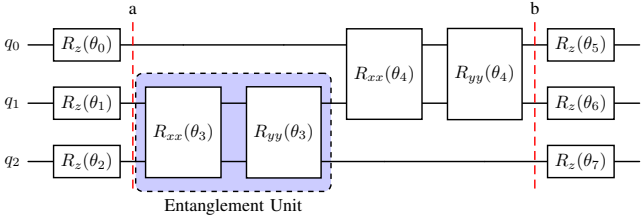
It is clear to see that $q_1$ and $q_2$ are entangled first, then $q_1$ and $q_0$ are entangled, which is in a reserved order of Fig.~\ref{fig:zzfeaturemap}.
\item \textit{Circular entanglement (\entalgCL):} Entanglement is applied between the first and the last qubits, after that linear entanglement is applied on all qubits. Taking the entanglement layer in Fig.~\ref{fig:excitationrefsim} as an example (i.e., between red dashed lines $a$ and $b$), \textit{excitation preserving} circuit here applies $R_{XX}$, $R_{YY}$, and $\mathit{CPhase}$ gates as an entanglement unit to entangle qubits.
\begin{figure}[!tb]
\centering
\resizebox{0.99\columnwidth}{!}{
\begin{quantikz}
\lstick{$q_0$} & \gate{R_z(\theta_0)} \slice{a} &\gate[3]{R_{xx}(\theta_3)}\gategroup[3,steps=3,style={dashed,rounded
corners,fill=blue!20, inner
xsep=2pt},background,label style={label
position=below,anchor=north,yshift=-0.2cm}]{{Entanglement Unit}}&\gate[3]{R_{xx}(\theta_3)}&\ctrl[wire style={"\theta_4"}]{2}& \gate[2]{R_{xx}(\theta_5)} & \gate[2]{R_{yy}(\theta_5)} & \ctrl[wire style={"\theta_6"}]{1} & & & \slice{b} &\gate{R_z(\theta_9)} & \\
\lstick{$q_1$} & \gate{R_z(\theta_1)} &&&& & & \control{} & \gate[2]{R_{xx}(\theta_7)} & \gate[2]{R_{yy}(\theta_7)} &\ctrl[wire style={"\theta_8"}]{1}& \gate{R_z(\theta_{10})}& \\
\lstick{$q_2$} & \gate{R_z(\theta_2)} &&&\control{}& & & & & &\control{} & \gate{R_z(\theta_{11})}&
\end{quantikz}
}
\caption{\textit{Excitation preserving} circuit for mode \textit{``fsim''} (\ansatzEPFS) with \textit{circular entanglement} (\entalgCL)}
\label{fig:excitationrefsim}
\end{figure}
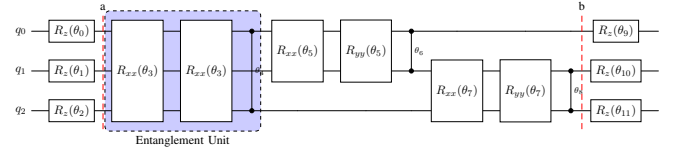
$q_0$ and $q_2$ are entangled before the linear entanglement applied on $q_0$ and $q_1$, as well as $q_1$ and $q_2$.
\item \textit{Shifted circular alternating entanglement (\entalgSCA):} It is implemented according to~\cite{sim2019expressibility}. The structure is similar to circular entanglement, except for the position of entanglement shifts positions in different repetitions. Also, the role control and target qubits are swapped in each repetition. Let's take the entanglement layer in Fig.~\ref{fig:efficientsca} as an example.
\begin{figure}[!tb]
\centering
\resizebox{0.99\columnwidth}{!}{
\begin{quantikz}
\lstick{$q_0$} & \gate{R_y(\theta_0)}&\gate{Z} \slice{a} &\gate{X} & \ctrl{1} & \slice{b} & \gate{R_y(\theta_3)} &\gate{Z} \slice{c} & &\ctrl{2}&\gate{X} \slice{d}& \gate{R_y(\theta_6)} &\gate{Z}& \\
\lstick{$q_1$} & \gate{R_y(\theta_1)}&\gate{Z} & & \gate{X}&\ctrl{1} & \gate{R_y(\theta_4)} &\gate{Z}&\gate{X}& &\ctrl{-1}& \gate{R_y(\theta_7)} &\gate{Z}& \\
\lstick{$q_2$} & \gate{R_y(\theta_2)}&\gate{Z} & \ctrl{-2} & &\gate{X} & \gate{R_y(\theta_5)} &\gate{Z}&\ctrl{-1}&\gate{X}& & \gate{R_y(\theta_8)} &\gate{Z}&
\end{quantikz}
}
\caption{\textit{Efficient SU2} circuit (\ansatzES) with \textit{sca entanglement} (\entalgSCA)}
\label{fig:efficientsca}
\end{figure}
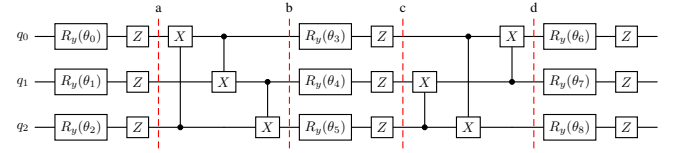
This example circuit contains two repetitions, and gates between dashed lines $a$, $b$, and $c$, $d$ are two entanglement layers. In the first layer, the implementation is the same as the circular entanglement. In the second layer, the entanglement among three qubits shifts positions with each other. In addition, the three pairs of control and target qubits are all swapped.
\end{compactitem}

Next, we introduce the quantum feature map and ansatz types used in this work.

\subsubsection{Quantum feature map}\label{sec:quantumFeatureMap}
In \ourApproach, we apply the \textit{ZZ feature map} to encode classical feature values of each request into quantum states. It is a quantum feature map that is specially designed for QML in a way that captures correlations between features with quantum entanglement~\cite{havlivcek2019supervised}.

To construct the feature map circuit for each request, $n$ qubits are required to map $n$ features. Consider Fig.~\ref{fig:zzfeaturemap} as an example. Suppose that three features are available, and we encode the three feature values $\boldsymbol{v}=\{v_0, v_1, v_2\}$ into the feature map circuit. As shown in the figure on the left side of the dashed line, each qubit is first applied with a Hardamard gate $H$ followed by a phase gate $P$, where the rotation angle of each phase gate is decided by one of the three feature values $v_i$. Next, entanglement operations are performed between pairs of qubits, as shown on the right side of the dashed line. We implement linear entanglement (i.e., \entalgLN) in the feature map where two $C_X$ gates and a phase gate are combined as a block to rotate the two applied qubits simultaneously by the angles of the corresponding feature values $v_i$ and $v_j$ in order to create superposition between $q_i$ and $q_j$. Note that $\varphi(v_i, vj)=(\pi - v_i)(\pi - v_j)$ in Fig.~\ref{fig:zzfeaturemap}. In this example, $q_0$ and $q_1$ are directly entangled, as are $q_1$ and $q_2$. 
This entanglement step captures non-linear relationships between the feature values.

\subsubsection{Ansatz types}
In this work, we select four widely used ansatz types to train the QNN model, which are {\it Real Amplitude}, {\it Pauli Two Design}, {\it Efficient SU2}, and {\it Excitation Preserving} circuits. The entanglement structures described are used in each ansatz implemented in this paper. The four types of ansatz comprise alternating rotation layers and entanglement layers. Typically, an additional rotation layer is added at the end of the ansatz.

\textbf{Real Amplitude} (\ansatzRA):
The Real Amplitudes circuit is used as the ansatz in both chemistry applications and classification circuits in quantum machine learning. Its rotation layer only consists of $R_Y$ gates, whose rotation angles are trainable parameters in the QNN circuit. These parameters are updated by the classical optimizer in a training process. The entanglement layer is composed of $C_X$ gates, whose structure can be flexible, as we can choose from any entanglement structure introduced before, though it does not include any trainable parameters. Fig.~\ref{fig:realfull} is an example of a Real Amplitude circuit with three qubits. $\theta_0$ to $\theta_5$ are trainable parameters.

\textbf{Pauli Two-Design} (\ansatzPTD): 
Pauli Two-Design circuits are two-design circuits~\cite{nakata2017unitary} frequently studied in QML~\cite{mcclean2018barren}. Pauli Two-Design circuit starts with an initial layer by applying $R_Y(\pi/4)$ gates on each qubit before the alternating rotation and entanglement layers (i.e., the left side of the dashed line $a$ in Fig.~\ref{fig:paulipair}). The rotation layer consists of rotation gates (i.e., $R_X$, $R_Y$, and $R_Z$ gates) that are uniformly sampled. The entanglement layer comprises $C_Z$ gates, which can only be arranged in a \textit{pairwise entanglement} (\entalgPW) structure. In the rotation layers shown in Fig.~\ref{fig:paulipair}, the standard rotation gates at all eight positions in the circuits are randomly selected.

\textbf{Excitation Preserving} (\ansatzEPIS and \ansatzEPFS):
Excitation preserving circuits are particularly important for applications in quantum chemistry and variational algorithms (e.g., QNN), which maintain the ratio between the $\ket{00}$, $\ket{01}+\ket{10}$, and $\ket{11}$ states to ensure the total number of excitations (i.e., qubits in the $\ket{1}$ state of being constant. The rotation layer is composed of $R_Z$ gates. The entanglement layer consists of $R_{XX}$ and $R_{YY}$ gates in ``iswap'' mode (\ansatzEPIS). In ``fsim'' mode (\ansatzEPFS), an extra $\mathit{CPhase}$ gate is applied to further entangle the two qubits. The entanglement structure is flexible. The rotation angles involved in the two layers are all trainable parameters. Fig.~\ref{fig:excitationreiswap} and Fig.~\ref{fig:excitationrefsim} show two examples respectively.

\textbf{Efficient SU(2) (\ansatzES):}
Efficient SU(2) circuits are ideal for classification tasks in quantum machine learning. SU(2) represents a class of quantum gates that includes both standard rotation gates and Pauli gates, which are used in the rotation layers. The number and type of gates applied in these layers can be selected and remain consistent across each repetition. If standard rotation gates are chosen, they contain trainable parameters. In contrast, if Pauli gates are used, they do not include any trainable parameters. Let's take Fig.~\ref{fig:efficientsca} as an example, which shows two repetitions. It applies two gates in each rotation layer, which are $R_{Y}$ and $Z$ gates. Thus, all rotation angles in $R_Y$ gates are the trainable parameters. Besides, the gates applied in the rotation layer are $C_X$ gates and the entanglement structure is flexible.



\section{Experiment Design}\label{sec:experimentDesign}

In this section, we describe the design of the experiment we conducted to assess \ourApproach. The code is available online at~\cite{SupplementaryMaterialQuantumCancerICST2025}. For IP protection, data and models cannot be shared.

\subsection{Research questions}\label{sec:rqs}

Our study has the goal of answering the following research questions (RQs):
\begin{compactitem}
\item[{\bf RQ1}:] \emph{How does \ourApproach compare to \evoclass, i.e., the classical approach that uses a classical machine learning model?}

The goal of this RQ is to assess whether applying QNNs is feasible for testing \caress, i.e., if it obtains performance results that are comparable to those of the classical machine learning approach \evoclass. 

\item[{\bf RQ2}:] \emph{What is the effect of each individual hyperparameter on the accuracy of \ourApproach?}

In this RQ, we analyze the effect of different hyperparameters individually on the model accuracy. We first investigate the significance of the four hyperparameters (i.e., number of features, number of repetitions, entanglement structure, and ansatz type) on the accuracy of \ourApproach. Next, for each individual hyperparameter, we recommend its optimal value. By identifying the best value for each hyperparameter, we can provide guidance for developers who want to tune their QNN models based on these key hyperparameters.

\item[{\bf RQ3}:] \emph{What is the effect of combined hyperparameters on the accuracy of \ourApproach?}

This RQ focuses on exploring the interaction between hyperparameters. Based on RQ2, by combining the hyperparameters having a higher contribution to the accuracy, we recommend the best-performing combinations of two and three hyperparameters. This recommendation allows developers to start with optimized combinations, reducing costs, while maintaining the flexibility to further tune their QNN model.
\end{compactitem}

\subsection{Experimental setup}\label{sec:experimentSetup}
\textbf{Experiment environment:} We have implemented \ourApproach on the ideal AerSimulator in the Qiskit framework. Experiments are run on Intel(R) Xeon(R) Platinum 8360Y CPU @ 2.40GHz.

\textbf{Datasets and features:} Due to the long QNN training time on a simulator, we randomly select a subset from the dataset created in the original paper~\cite{isaku2023cost}, using 1000 instances for training and 500 instances for testing. Additionally, to train the QNN, out of 56 original features we select subsets of the most important ones from the dataset, based on their importance scores in the original paper.

\textbf{QNN hyperparameters:} We train \ourApproach using different combinations \comb = \combTuple{\featNum}{\repNum}{\entalg}{\ansatz} of its hyperparameters: number of features (\featNum), number of repetitions (\repNum), entanglement structure (\entalg), and ansatz type (\ansatz). We identify with \ourApproachComb{\comb} the setting of \ourApproach with \comb. Specifically, we experiment with $\featNum \in \{4,$ $5,$ $6,$ $7,$ $8\}$, $\repNum \in \{1,$ $3,$ $5,$ $7\}$, $\entalg \in \{\entalgFL,$ $\entalgLN,$ $\entalgRL,$ $\entalgPW,$ $\entalgCL,$ $\entalgSCA\}$, and $\ansatz \in \{\ansatzRA,$ $\ansatzES,$ $\ansatzEPFS,$ $\ansatzEPIS,$ $\ansatzPTD\}$. For the ansatz types \ansatzRA, \ansatzEPFS, \ansatzEPIS, and \ansatzPTD, we conduct experiments by pairing each with the six different entanglement structures \entalg. The ansatz \ansatzES, instead, has a fixed entanglement structure \entalgPW. Thus, we build a total of 25 types of QNN circuits. Additionally, when setting \featNum, we select the top $4$, $5$, $6$, $7$, and $8$ most important features from the original dataset to train the QNN. 
We further implement each QNN circuit with all 4 different numbers of repetitions (i.e., 1, 3, 5, 7). Thus, overall, we construct and train 500 QNN models.

\textbf{Baseline:} We train the classical ML model \evoclass with the optimal setting from the paper~\cite{isaku2023cost}, using the same dataset as \ourApproach but with the original 56 features.

\textbf{Parameter settings:} In \ourApproach, the classical optimizer we select is \textit{Constrained Optimization By Linear Approximation} optimizer (COBYLA)~\cite{powell2007view} with 400 iterations, which is commonly used in the literature, combined with squared error as the loss function. To tackle the non-determinism of the training process, we repeat the training for \evoclass and each QNN model 10 times.

\subsection{Statistical tests and evaluation metrics}\label{sec:evaluationMetrics}

In order to answer RQ1, we first compute the average accuracy $\averageAccuracy_{\comb}$ (across the ten runs) of each version of \ourApproachComb{\comb}. Then, we sort all the versions of the approach, and we experimentally select the combination \bestComb that has the highest $\averageAccuracy_{\bestComb}$. Finally, we compare the $\averageAccuracy_{\bestComb}$ with the accuracy of the classical model $\averageAccuracy_{\evoclass}$.

To answer RQ2, we apply the analysis of variance (ANOVA) on all four hyperparameters (i.e., \featNum, \repNum, \ansatz, \entalg) to evaluate their contribution to the accuracy of \ourApproach. If the $p$-value for a hyperparameter is smaller than 0.05, the variations in this hyperparameter can significantly affect the accuracy. Otherwise, there is no significant effect. On the other hand, a higher $F$ statistics for a hyperparameter represents a greater influence on the accuracy of \ourApproach. Then, for each hyperparameter with significant effect, we identify the optimal value by ranking the average accuracy $\averageAccuracy_{\hparm}$ (across all other hyperparameters among the ten runs) of each value. 

To answer RQ3, we calculate the average accuracy $\averageAccuracy_{\comb}$ of pairwise and three-way combinations (across all other hyperparameters with ten runs) of hyperparameters with the top two contribution and top three contribution respectively, then we rank them to find the optimal combination values (i.e., \bestCombTwo and \bestCombThree) that achieve the highest accuracy (i.e., $\averageAccuracy_{\bestCombTwo}$ and $\averageAccuracy_{\bestCombThree}$).

\section{Experimental results}\label{sec:results}

\subsection{RQ1 -- How does \ourApproach compare to \evoclass, i.e., the classical approach that uses a classical machine learning model?}

In this RQ, we want to assess whether the application of \ourApproach is feasible in practice, i.e., if it can provide results comparable to those of \evoclass that is based on a classical ML model, i.e., random forest. 

Notably, as explained in Sect.~\ref{sec:experimentSetup}, due to the limited quantum resources available, we cannot use all the information in the dataset to train a QNN model, but we can only rely on a small number of features (i.e., $\featNum \in \{4,$ $5,$ $6,$ $7,$ $8\}$ among all 56 features). If this proves to be effective, it will justify further investigation into QNN hyperparameters for future performance improvement of the application on \caress even with current quantum resources.

We assess the feasibility of \ourApproach by experimentally selecting the combination \bestComb of hyperparameters \combTuple{\featNum}{\repNum}{\entalg}{\ansatz} that provides the best average accuracy $\averageAccuracy_{\bestComb}$. We observe that such combination is \bestComb = \combTuple{5}{7}{\entalgSCA}{\ansatzEPIS}, achieving $\averageAccuracy_{\bestComb} = 0.921$. The average accuracy (among 10 runs) of the classical model \evoclass trained on 56 features, instead, is $\averageAccuracy_{\evoclass} = 0.946$. The two accuracy values are both higher than 0.90, and the difference is smaller than 0.03, which is relatively small and shows that a QNN can closely approximate the performance of a well-established classical method, which is a significant result in the context of early applications of quantum approaches, even with limited quantum computation resources. Therefore, we can conclude that \ourApproach is a valid alternative to \evoclass as it provides comparable accuracy results.

We believe that, with more investigation of QML and advancement of quantum hardware, it would be possible to overcome obstacles such as hardware limitation, expand ansatz selection and scalability (not only two-local circuits), develop more effeicient encoding techniques, and so on; these points will allow us to fill the small gap that currently exists in the accuracy results. In addition, the inherent high computing speed of quantum computing will show the performance advantage of QNN over classical ML models in the future.

\begin{tcolorbox}[size=title, colframe=white, width=1\linewidth,
breakable,
colback=gray!20]
\textbf{Answer to RQ1}:
\ourApproach demonstrated to be feasible to be applied to test \caress using the limited number of features and training on constrained quantum computing resources. It has great potential to be an alternative to \evoclass in the future.
\end{tcolorbox}

\subsection{RQ2 -- What is the effect of each individual hyperparameter on the accuracy of \ourApproach?}
In this RQ, we aim to analyze the individual impact of each hyperparameter (i.e., \featNum, \repNum, \entalg, and \ansatz) on the performance of \ourApproach. First of all, we identify the key parameters that can significantly affect the accuracy. We apply the statistical test ANOVA on each hyperparameter based on the performance of each QNN model with 10 runs. Table~\ref{tab:anovaRQ2} reports the results in terms of $p$-values and $F$ statistics.
\begin{table}[!tb]
\centering
\caption{RQ2 -- Results of ANOVA}
\label{tab:anovaRQ2}
\resizebox{0.99\columnwidth}{!}{
\begin{tabular}{ccccc}
\toprule
& \featNum & \repNum & \entalg & \ansatz \\
\midrule
$F$ statistics & 1480.76 & 303.35 & 16.21 & 0.03 \\
$p$-value & $6.46\times 10^{-150}$ & $3.44 \times 10^{-53}$ & $8.49 \times 10^{-15}$ & $9.98 \times 10^{-1}$ \\
\bottomrule
\end{tabular}
}
\end{table}
We observe that $p$-values of \entalg, \featNum, and \repNum are all much lower than 0.05, indicating that all three hyperparameters have a significant effect on the accuracy of the QNN model. According to the $F$ statistics, we can identify that \featNum has the largest effect on the accuracy, followed by \repNum and \entalg.

Instead, the $p$-value of \ansatz is higher than 0.05, indicating that we cannot observe a clear trend in accuracy when changing \ansatz alone. This implies that any noticeable impact might only emerge when \ansatz is combined with other hyperparameters. Consequently, in the context of this paper's case study, changing the ansatz type in a QNN model does not significantly affect performance, potentially due to the characteristics of the dataset and the classification task. Nonetheless, in other case studies or different tasks, \ansatz could prove to be a critical factor. 

Regarding the three hyperparameters that have a significant impact, we can analyze how the setting of each hyperparameter affects the performance. Based on this analysis, we recommend the optimal value for each parameter so that developers of QNN can have clear starting points for optimizing their own models. In addition, they can focus on individual parameter tuning before considering combinations, to reduce the complexity of the initial experiments. 

\begin{figure*}[!tb]
\centering
\begin{subfigure}{.32\textwidth}
\centering
\includegraphics[width=1\linewidth]{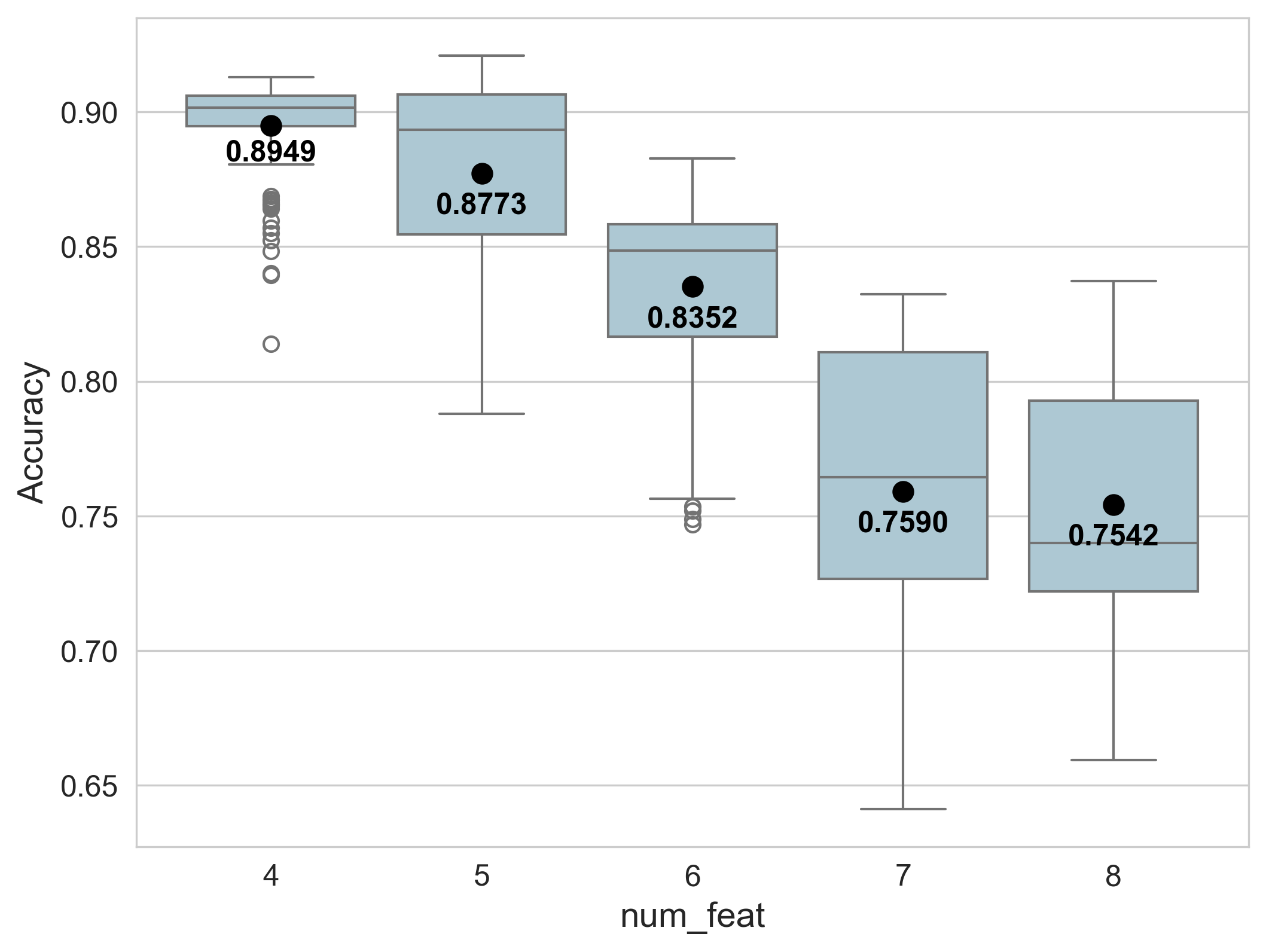}
\caption{Number of features (\featNum)}
\label{fig:rq2boxplotNumFeat}
\end{subfigure}%
\begin{subfigure}{.32\textwidth}
\centering
\includegraphics[width=1\linewidth]{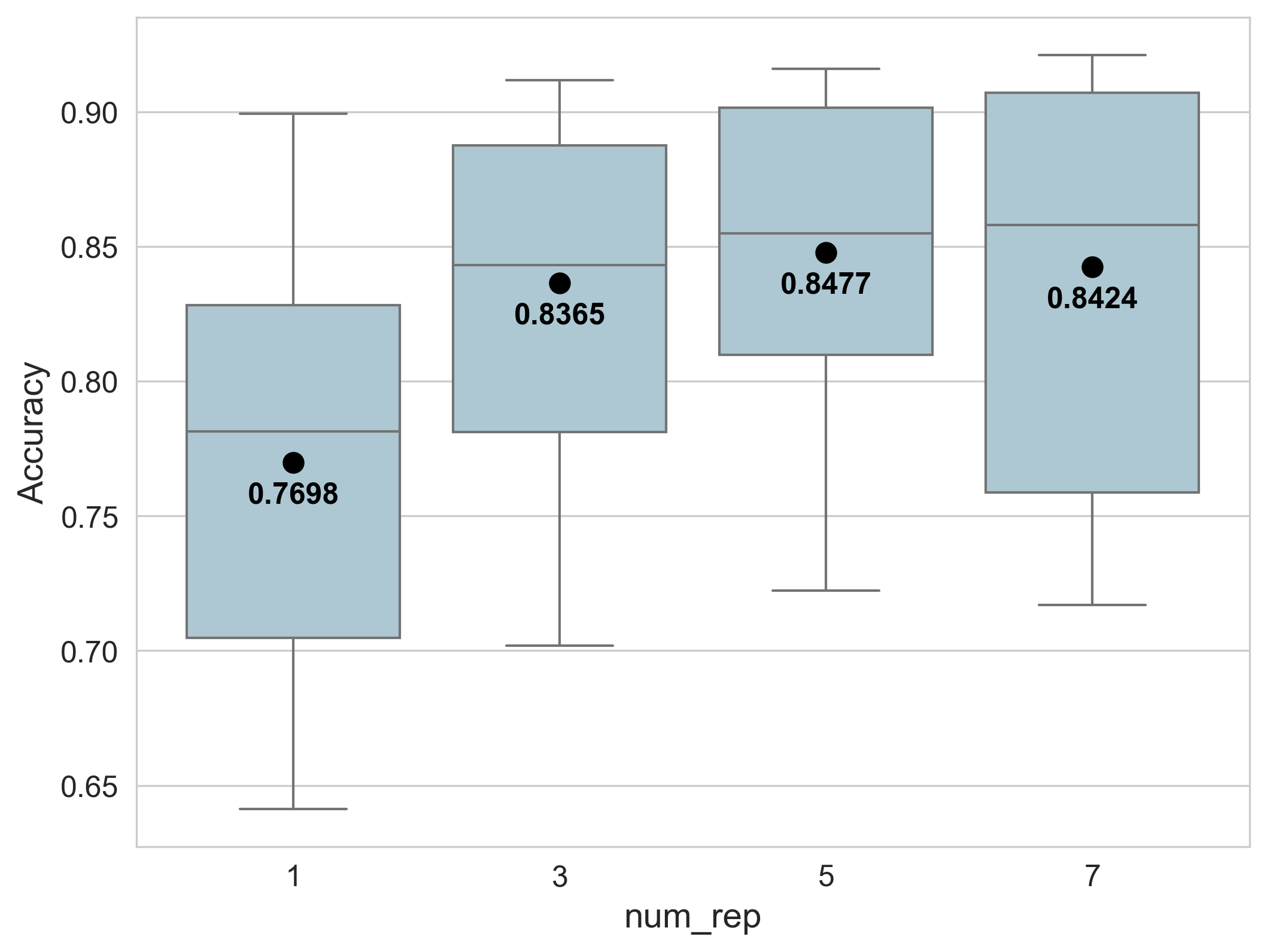}
\caption{Number of repetitions (\repNum)}
\label{fig:rq2boxplotNumRep}
\end{subfigure}
\begin{subfigure}{.32\textwidth}
\centering
\includegraphics[width=1\linewidth]{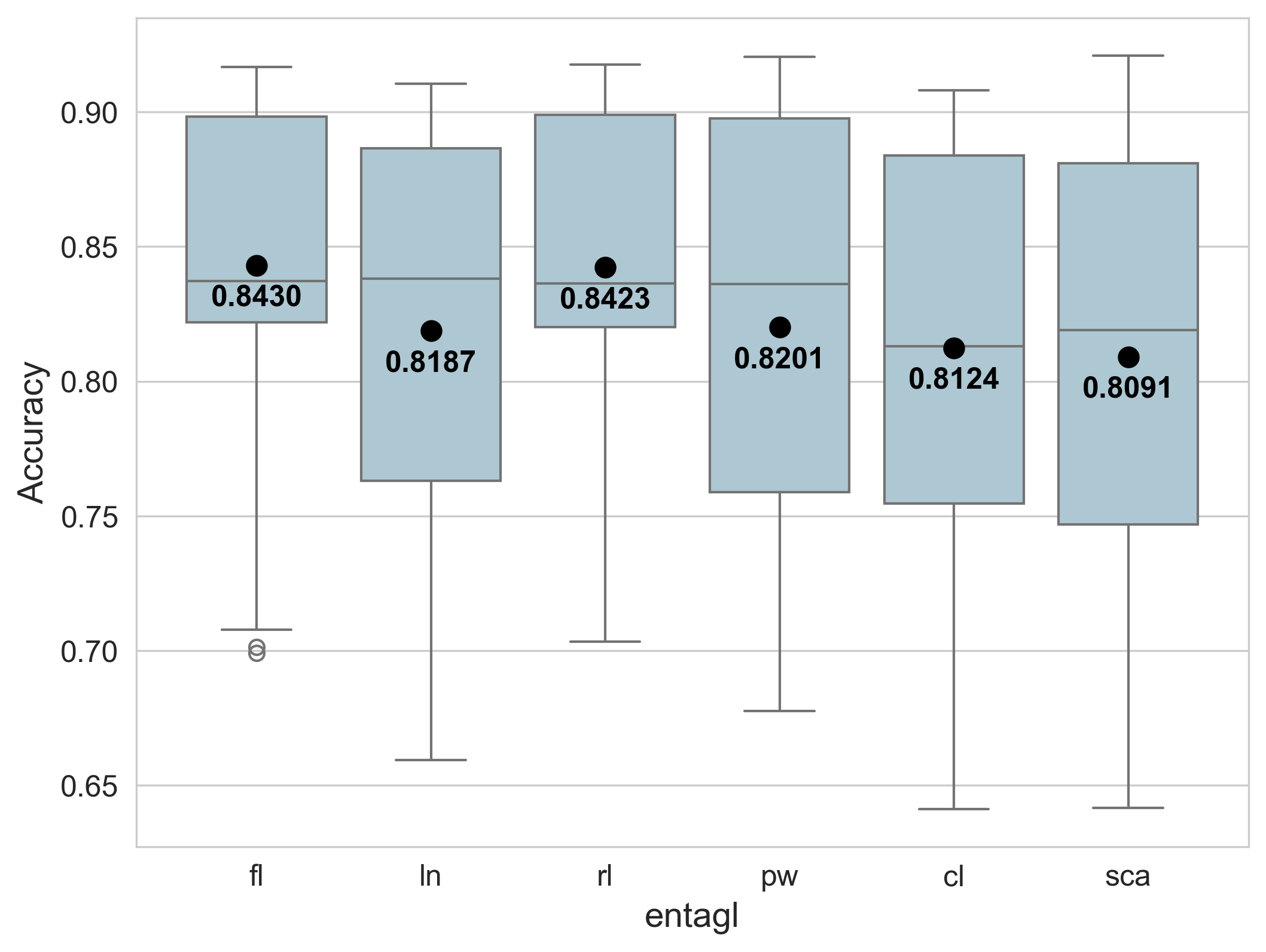}
\caption{Entanglement (\entalg)}
\label{fig:rq2boxplotEntalg}
\end{subfigure}
\caption{RQ2 -- Effect of individual hyperparameter on accuracy of \ourApproach. Black dots $\bullet$ represent average values.}
\label{fig:rq2boxplots}
\end{figure*}

In Fig.~\ref{fig:rq2boxplots}, we report the boxplots of accuracy values for the three hyperparameters having a significant impact on accuracy.
Each boxplot contains data points of all combinations with a specific feature value. In Fig.~\ref{fig:rq2boxplotNumFeat}, the trend is clear: \featNum with lower values (i.e., 4 and 5) tend to achieve higher accuracy, followed by $\featNum=6$. \featNum of 7 and 8 achieve the worse accuracy. Additionally, $\featNum=4$ achieves the optimal performance among the 5 values with $\averageAccuracy_{\featNum=4}=0.895$. There are two possible explanations for these results.

First, QNN with higher feature numbers (i.e., number of qubits) may require more iterations in the classical optimizer, along with more complex feature mapping and ansatz circuits. However, in this work, considering the limits of current quantum computers, we are investigating circuits with gates involving at most two qubits, without considering gates involving three qubits (which could provide better results. Thus, in the future, when implementing other ansatz types, and feature mapping, the performance with higher qubits may improve.

Second, it is also possible that the sensitivity of QNN is higher than that of classical approaches. Since we select the top important features for training each time, using less important features in the QNN model (e.g., the additional four features used in $\featNum=8$ comparing $\featNum=4$ ) could potentially reduce performance instead.

Regarding the number of repetitions \repNum (see Fig.~\ref{fig:rq2boxplotNumRep}), we can see a clear increasing trend when applying QNN with higher \repNum values. 
This could be due to the fact that increasing the number of repetitions in a QNN circuit will increase the depth of the circuit, and, consequently, the number of trainable parameters; this enables the model to capture complex data patterns and handle more intricate relationships between the input features and output.
However, due to the effect of low performance of \ourApproach with 7 and 8 features when $\repNum=7$ (which will be introduced in RQ3), the optimal value is $\repNum=5$ with $\averageAccuracy_{\repNum=5}=0.848$. 

Regarding \entalg (see Fig.~\ref{fig:rq2boxplotEntalg}), the optimal value is $\entalg=\entalgFL$, which is probably because it is the only entanglement structure in which any two qubits are directly entangled between each other when there are 4 or more qubits in the circuit. This means that higher entanglement may increase the ability to capture the data pattern.






\begin{tcolorbox}[size=title, colframe=white, width=1\linewidth,
breakable,
colback=gray!20]
\textbf{Answer to RQ2}:
\featNum has the strongest influence on the accuracy of \ourApproach, followed by \repNum and \entalg, while \ansatz shows no significant effect. To provide developers with optimal starting points and minimize extensive tuning, we recommend the best values for each hyperparameter with a positive impact: $\featNum=4$, $\repNum=5$, and $\entalg=\entalgFL$.
\end{tcolorbox}

\subsection{RQ3: What is the effect of combined hyperparameters on the accuracy of \ourApproach?}
We investigate the interaction among hyperparameters since some hyperparameters might interact in ways that enhance or diminish each other's effect, which cannot be captured by observing them individually.
Thus, we analyze the hyperparameter interactions by observing the impact of parameter combinations on the accuracy of QNN.

\begin{figure}[!tb]
\centering
\includegraphics[width=0.99\columnwidth]{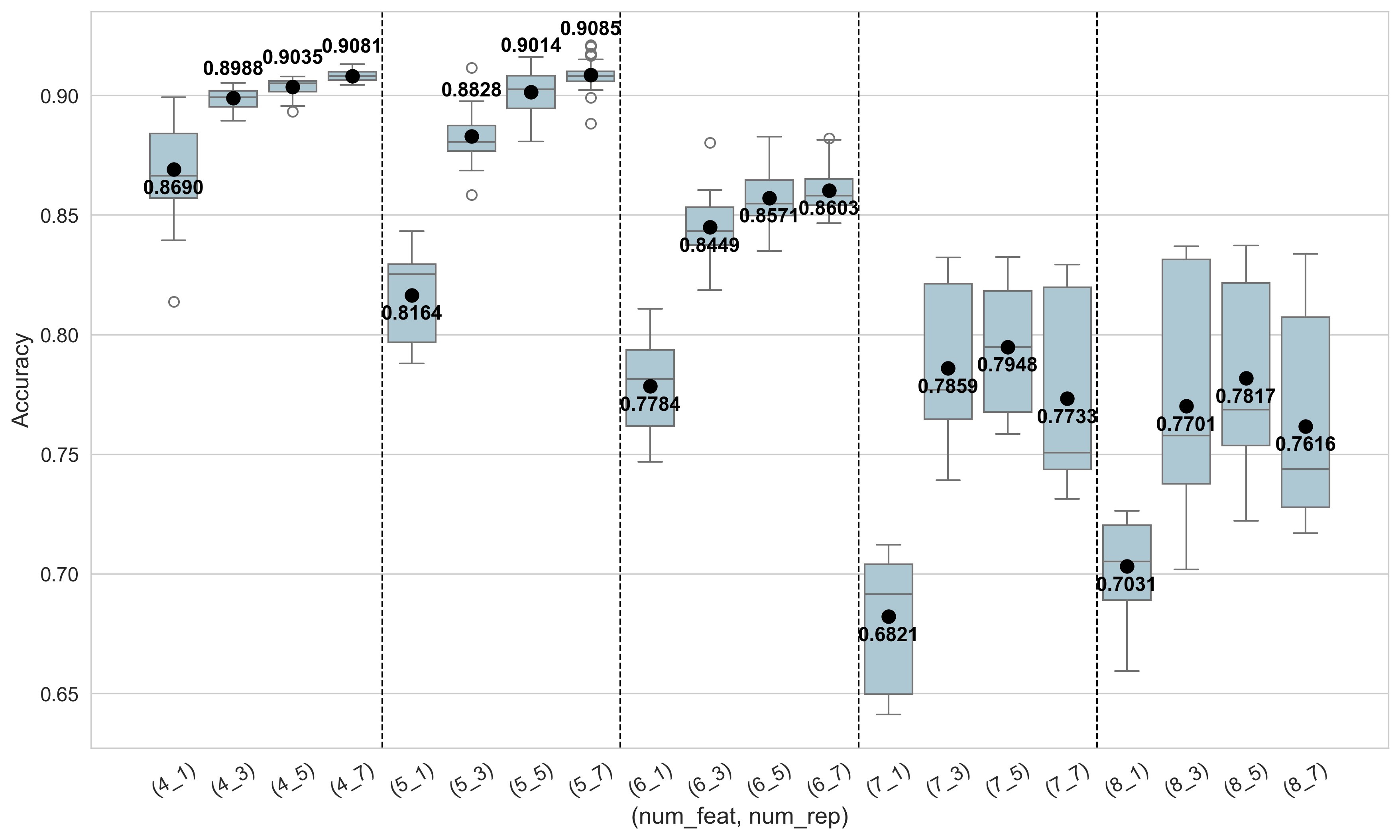}
\caption{RQ3 -- Effect of hyperparameter pairs (\featNum, \repNum) on the accuracy of \ourApproach. Black dots $\bullet$ represent average values.}
\label{fig:rq3}
\end{figure}

First, we analyze the effect of the hyperparameter pairs with the top two contributions (i.e., \featNum and \repNum). Fig.~\ref{fig:rq3} shows the boxplot of the accuracy across various values combination.
%
We observe that, with $\featNum \in \{4, 5\}$ and $\repNum \in \{3, 5, 7\}$, the accuracy values are consistently higher than 0.85. Furthermore, with $\featNum \in \{4,$ $5,$ $6\}$, accuracy tends to improve when \repNum increases. Also, when $\featNum \in \{4,$ $5,$ $6\}$ and $\repNum \in \{3, 5, 7\}$, the accuracy variation is relatively smaller than other combinations. However, it is not the same case when \featNum equals $7$ and $8$, where no clear increasing pattern can be observed across \repNum values of $3$, $5$, $7$. The average accuracy values of $\repNum=5$ are the highest in both cases, followed by $\repNum=3,$ $7$, and $1$. Additionally, when $\featNum \in \{7,$ $8\}$ and $\repNum \in \{3,$ $5,$ $7\}$, the accuracy variation is generally larger than other combinations. Notably, when $\repNum=1$, with the same \featNum, the accuracy is much lower than any other \repNum values. By ranking the average values $\averageAccuracy_\comb$, we find the optimal combination is \bestCombTwo = \combTupleTwo{5}{7}, who achieves $\averageAccuracy_{\bestCombTwo} = 0.909$. 

Next, we analyze the effect of the hyperparameter combination consisting of the three with significant contributions (i.e., \featNum, \repNum, and \entalg). We calculate \averageAccuracy of \combTupleThree{\featNum}{\repNum}{\entalg}. Following the above conclusions, we find that \bestCombTwo together with the optimal \entalg (i.e., \entalgFL) achieves the highest accuracy $\averageAccuracy_{\bestCombThree} = 0.912$. Thus, we can conclude that \bestCombThree = \combTupleThree{5}{7}{\entalgFL}.








\begin{tcolorbox}[size=title, colframe=white, width=1\linewidth,
breakable,
colback=gray!20]
\textbf{Answer to RQ3}:
We recommend the best combination values of \combTupleTwo{\featNum}{\repNum} and \combTupleThree{\featNum}{\repNum}{\entalg}. Results show that QNN performs best with $\featNum=5$ and $\repNum=7$. Furthermore, with $\entalg=\entalgFL$, it reached the highest average accuracy of 0.912.
\end{tcolorbox}




\section{Discussion}\label{sec:discussion}
In this section, we present the key lessons learned that are valuable for researchers and practitioners.

\paragraph*{\bf Recommendations for Future Development}
In this paper, we evaluate the QNN models across various configurations by changing the four hyperparameters (i.e., \featNum, \repNum, \entalg, and \ansatz). We first investigate the individual impact of each hyperparameter on the performance of the QNN. We find that three hyperparameters (i.e., \featNum, \repNum, \entalg) can significantly affect the performance, where \featNum has the highest impact, followed by \repNum and \entalg. Next, we show how the specific settings of each hyperparameter, as well as interactions among those hyperparameters, affect the performance. Additionally, we present the optimal values for individual hyperparameters and their combinations.

By approaching the problem stepwise (i.e., first analyzing individual effects, then interactions), we offer practitioners a practical, structured way to tune QNNs, reducing the need for brute-force parameter testing across all combinations. Practitioners can have a clear starting point from tuning one specific hyperparameter and then moving to combinations. The findings of this work can benefit developers of QNNs, as they will have both general guidelines and specific recommendations for building well-performing models.




\paragraph*{\bf Potential of Quantum Neural Networks}
In this paper, we demonstrate that QNNs, even with a simple QNN model, can achieve performance similar to that of the classical approach \evoclass. Specifically, in \ourApproach, we use at most 8 features to train the model; instead, \evoclass uses 56 features for training. This shows that quantum circuits can encode low-dimension classical data into high-dimensional data to enhance the model's prediction ability.

Moreover, the trainable parameters used in \ourApproach are far fewer than those used in \evoclass. Specifically, the hyperparameter combination with the best performance \combTuple{5}{7}{\entalgSCA}{\ansatzEPIS} contains only 75 parameters. 
However, in \evoclass, the number of trainable parameters in the classical machine learning model is consistently higher than 16,000, which is over 200 times that of the QNN. We did a small experiment by comparing the two models with a closer number of trainable parameters, in order to get more insight into the potential advantage of QNN over the classical approach. We reduced the size of the random forest inside \evoclass as follows: $\texttt{n\_estimator}=1$, $\texttt{max\_feature}=\texttt{None}$, $\texttt{min\_samples\_split}=2$, $\texttt{min\_samples\_leaf}=1$. In order to keep the overall robustness of the model, we kept $\texttt{max\_depth}=2$. As a result, we reduced the number of trainable parameters to a range of 275 to 325 parameters. We trained the model 10 times. The average accuracy we achieved is $0.894$ among 10 runs, which is lower than the best configurations achieved in \ourApproach. It indicates that QNN models can learn more efficiently than the classical random forest with fewer trainable parameters since unique features of quantum computing (e.g., entanglement and superposition) can capture data correlations with fewer trainable parameters. Such efficiency can reduce over-fitting and improve generalization, especially in scenarios where it is difficult to collect large amounts of training data. It also saves time and energy for training on the real hardware.

Given that we use simple QNN models, we expect even better performance in future work. For example, hybrid quantum neural networks can be considered to enhance the performance~\cite{blance2021quantum}.

\paragraph*{\bf Generalizability to Other Software Engineering Problems}
In this paper, we present the application of QNN on a binary classification problem aiming at detecting invalid requests generated by \evomaster when testing \caress. However, the implementation of QNN can be adapted to other classification problems in the software engineering domain, such as software defect prediction, as our work demonstrates the great potential of QNN in practical software engineering classification tasks. The practitioners can follow our recommendation to build effective QNN with optimized performance while saving computational resources. Additionally, this work opens avenues for further research to enhance QNN performance and explore the applicability of other QML models for different software engineering tasks, such as requirements optimization, test prioritization, and test generation.

\paragraph*{\bf Limitations}
Given that this work is a new exploration of applying QNN to classification problems in software engineering, there are still some limitations. First, the implementation of the QNN models is on an ideal simulator in the Qiskit framework, without considering the effect of hardware noise. Such noise is unavoidable on current quantum computers. One of our main future works will be to evaluate the noise effect and integrate noise handling approaches~\cite{quantumNoiseTSE2024,errorMitigationQuantCompFSE2024}.

In addition, due to the slow QNN training process on the simulator, we only consider simple two-local circuits (i.e., feature mapping and ansatz) to build the QNN model. However, in the future, with more accessible quantum computers, more complex QNN models can be built to enhance performance.

\section{Threats to Validity} \label{sec:threats}
In \ourApproach, we train the QNN using a limited number of features and a smaller dataset size due to the long training time on the quantum simulator. Despite these constraints, results show that QNN can still achieve competent results. Similarly, \evoclass trained on the subset dataset performs comparably to the results in the original study~\cite{isaku2023cost}. Further, we select, $\repNum\in \{1,$ $3,$ $5,$ $7\}$ since there is no significant performance difference among consecutive values. Also, we cannot see significant improvement with higher values (e.g., $\repNum=9$). To reduce the randomness of QNN and \evoclass, we repeat 10 times and calculate the average accuracy value.

\section{Related work}\label{sec:related}
In recent years, advancements in quantum computing have driven increasing interest among researchers in investigating its potential to enhance machine learning algorithms, referred to as quantum machine learning (QML). Within QML, quantum neural network (QNN) is a widely used algorithm, which has been applied in various fields~\cite{gujju2024quantum, abbas2021power}. In the high-energy physics field, for example, Blance et al.~\cite{blance2021quantum} constructed a hybrid QNN to distinguish rare signals from a large number of standard model background events. Cugini et al.~\cite{cugini2023comparing} systematically compared the performance of QNN with classical deep learning algorithms on the signal and background classification problem. Results show that current quantum computers have reached performance comparable to classical deep learning algorithms. In finance, Thakkar et al.~\cite{thakkar2024improved} applied QNNs for credit risk assessment with far fewer parameters and higher accuracy than the classical counterpart. Zoufal et al.~\cite{zoufal2019quantum} applied Quantum Generative Adversarial Networks (qGAN) to enhance the efficiency of loading classical data into quantum states and demonstrated the performance of facilitating financial derivative pricing. In healthcare, QNN was applied for medical image classification~\cite{mathur2021medical}. Innan et al.~\cite{innan2024fedqnn} introduced the Federated Quantum Neural Network (FedQNN) framework and applied it to classification problems in a breast cancer dataset and a custom synthetic DNA dataset, proving high accuracy and suitability of QML for various tasks. In contrast to these works, we explore QNNs' potential for classification in software testing by integrating QNN with \evomaster, a test case generation tool, to support continuous testing in \caress, a real-world software system.

Building on this growing enthusiasm for quantum computing applications, several works have been proposed to tackle realistic problems in software testing. For example, Miranskyy~\cite{miranskyy2022using} implements the Grover's search algorithm to speed up dynamic testing in classical software. Quantum extreme learning machines~\cite{quantumMLelevatorsFSE2024,muqeet2024assessing} have demonstrated the potential of solving waiting time prediction problems in the context of regression testing of an industrial elevator. In addition, quantum optimization problems such as quantum annealing and the quantum approximate optimization algorithm have been applied for test case optimization problems~\cite{testMinQuantumAnnTOSEM2024,testMinqaoaTSE2024}. In this work, we employ QNN for solving a classification problem in the context of testing \caress at the Cancer Registry of Norway, a new application area of QNNs.




\section{Conclusion}\label{sec:conclusions}
We propose \ourApproach to investigate the feasibility of applying QNN for testing a real-world industrial system, the Cancer Registration Support System (\caress). We use \ourApproach to predict the invalidity of requests generated by \evomaster to avoid unnecessary testing costs. Results show that \ourApproach achieves comparable performance with the classical machine learning approach \evoclass. We also investigate how different QNN configurations affect performance and provide recommendations for future developers of QNN according to our empirical findings. In the future, we plan to improve QNN by investigating more ansatz types and feature maps, and exploring the potential of hybrid QNN. In addition, we will consider the effect of hardware noise on our approach.

\bibliographystyle{IEEEtran}
\bibliography{biblio}
\end{document}